\documentclass[usenatbib]{mn2e}

\usepackage{graphicx}
\usepackage{time}

\newcommand{\apj}{\mbox{\normalfont ApJ}}
\newcommand{\apjs}{\mbox{\normalfont ApJS}}
\newcommand{\apjl}{\mbox{\normalfont ApJL}}
\newcommand{\aj}{\mbox{\normalfont AJ}}
\newcommand{\mnras}{\mbox{\normalfont MNRAS}}
\newcommand{\aap}{\mbox{\normalfont A\&A}}

\newcommand{\nat}{\mbox{\normalfont Nature}}

\title[The BBs of the MW Halo]{On the Use of the Number Count of Blue Horizontal--Branch Stars to Infer the Dominant Building Blocks of the Milky Way Halo}
\author[C. Chung, Y.-W. Lee, and M. Pasquato]{Chul Chung$^{1}$\thanks{E-mail: chung@galaxy.yonsei.ac.kr}, Young-Wook Lee$^{1}$\thanks{E-mail: ywlee2@yonsei.ac.kr} and Mario Pasquato$^{1,2}$\\
$^{1}$Department of Astronomy \& Center for Galaxy Evolution Research, Yonsei University, Seoul 120-749, Republic of Korea\\
$^{2}$Yonsei University Observatory, Seoul 120-749, Republic of Korea}
\begin{document}

\date{Accepted 2015 October 16. Received 2015 October 13; in original form 2015 June 3}

\pagerange{\pageref{firstpage}--\pageref{lastpage}} \pubyear{2015}

\maketitle

\label{firstpage}

\begin{abstract}

The formation of the Milky Way stellar halo is thought to be the result of merging and accretion of building blocks such as dwarf galaxies and massive globular clusters.
Recently, \citet{2015MNRAS.448L..77D} suggested that the Milky Way outer halo formed mostly from big building blocks, such as dwarf spheroidal galaxies, based on the similar number ratio of blue straggler (BS) stars to blue horizontal-branch (BHB) stars.
Here we demonstrate, however, that this result is seriously biased by not taking into detailed consideration on the formation mechanism of BHB stars from helium enhanced second-generation population.
In particular, the high BS-to-BHB ratio observed in the outer halo fields is most likely due to a small number of BHB stars provided by GCs rather than to a large number of BS stars.
{This is supported by our dynamical evolution model of GCs which shows preferential removal of first generation stars in GCs.
Moreover, there are a sufficient number of outer halo GCs which show very high BS-to-BHB ratio.}
Therefore, the BS-to-BHB number ratio is not a good indicator to use in arguing that more massive dwarf galaxies are the main building blocks of the Milky Way outer halo.
Several lines of evidence still suggest that GCs can contribute a significant fraction of the outer halo stars.

\end{abstract}


\section{Introduction}

The recent study of \citet{2015MNRAS.448L..77D} reports measurements of the number ratio of blue stragglers (BS) to blue horizontal branch (BHB) stars in the Milky Way globular clusters (GCs) and dwarf satellite galaxies.
They found that the number ratio of the BS-to-BHB in luminous dwarf galaxies is similar to that of outer halo field stars (8.5~kpc ${\rm \leq R_{GC} \leq}$ 45~kpc), which led them to claim that the formation of the Milky Way halo favored more massive building blocks (i.e., luminous dwarf spheroidal galaxies) rather than GCs.
Their procedure involves a census of BHB stars in the color-magnitude diagram (CMD), and the purpose of this Letter is to show that this led them to reach a biased result.

\begin{figure*}
\includegraphics[angle=-90,scale=1.0]{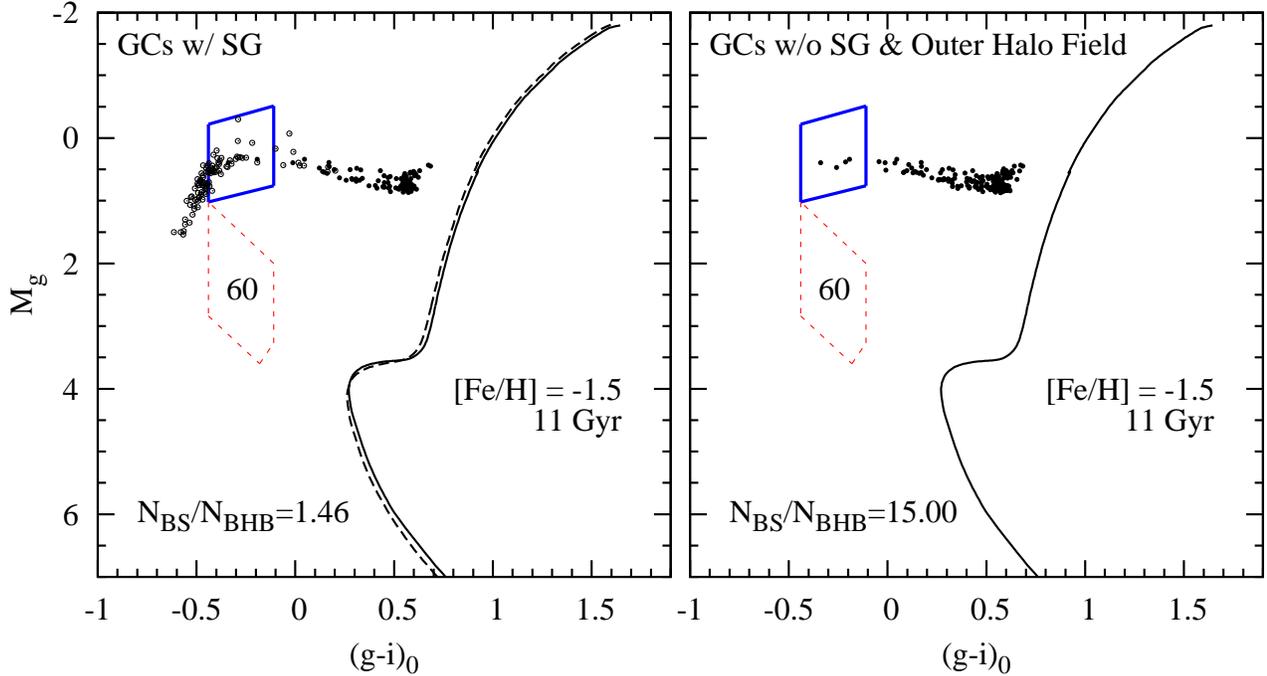}
\caption[]{
Synthetic color-magnitude diagrams (CMDs) and their ${\rm N_{BS}/N_{BHB}}$ ratios for old stellar systems with and without helium enhanced SG stars.
Solid and dashed lines are model isochrones for the FG (${\rm Y=0.23}$) and SG populations (${\rm Y=0.28}$), respectively, and the filled and open circles are corresponding HB stars.
In our model, the field stars in the outer halo are similar to the GCs without SG stars (see the text).
The selection boxes for BHB (blue box) and BS stars (red dashed box) are from \citet{2015MNRAS.448L..77D}.
The adopted ${\rm N_{BS}/N_{HB}}$ ratio is 0.3 and the total number of HB stars is 200.
}
\label{f1}
\end{figure*}

\begin{figure*}
\includegraphics[angle=-90,scale=0.6]{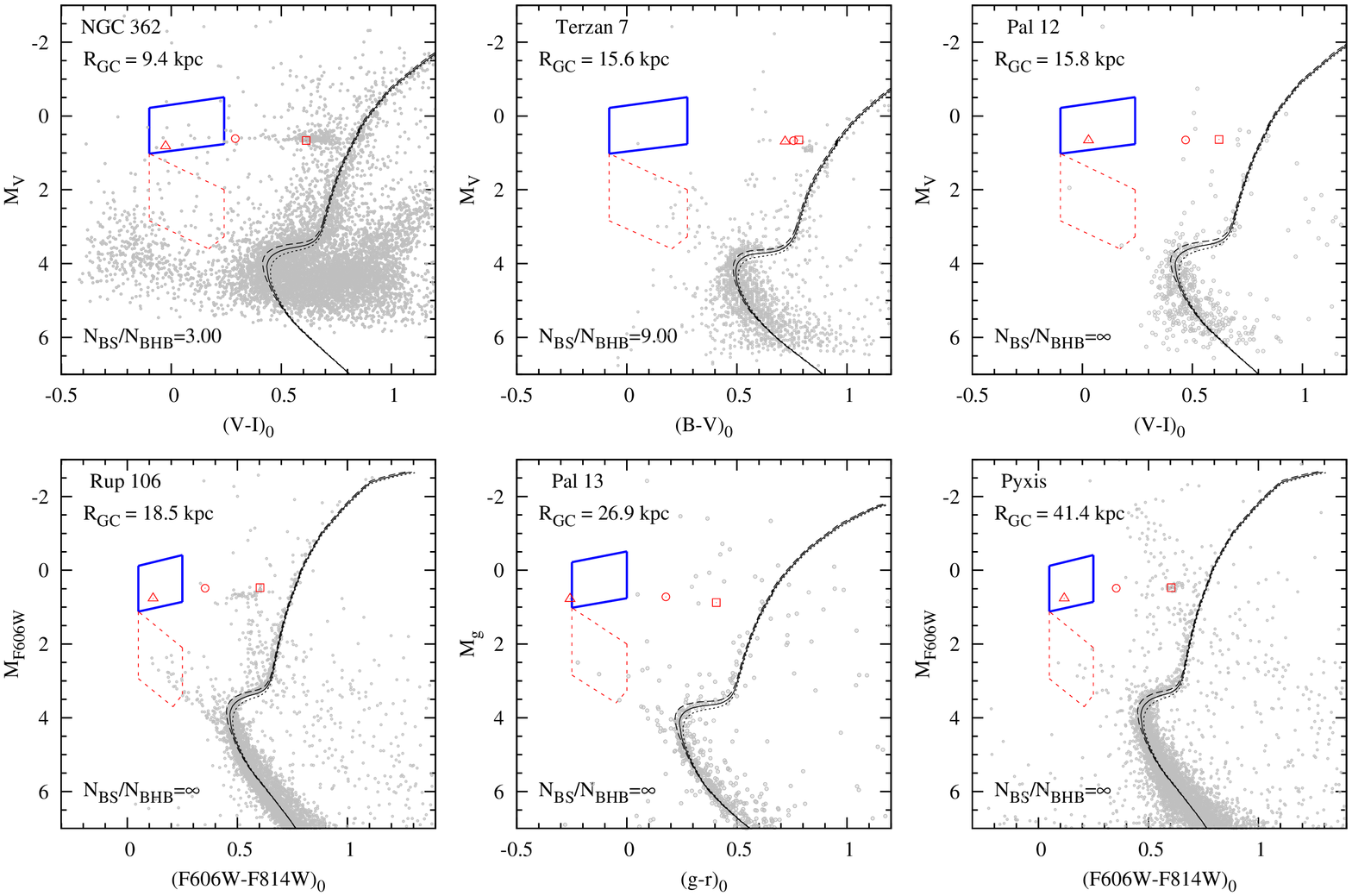}
\caption[]{
{Example of GC CMDs at $8 \sim 45$~kpc with no or small number of BHB stars.
Sources of the data and the result of isochrone fittings are summarized in Table~\ref{t1}.
The selected boxes in each CMD are modified to classify the same stars in Figure~\ref{f1}.
Black long dashed, solid, and dotted lines are isochrones of age at 11, 12, and 13~Gyr, respectively.
Red open square, circle, and triangle are provided to indicate the mean position of HB distributions at 11, 12, and 13~Gyr, respectively.
The ${\rm N_{BS}/N_{BHB}}$ of selected CMDs increases as the number of BHB stars decreases.}
}
\label{f2}
\end{figure*}

\section{Blue HB Stars from Helium Enhanced Second-Generation Stars}

The HB morphology is sensitive to parameters such as metallicity, age, helium, and CNO abundances \citep{1994ApJ...423..248L, 2008MNRAS.390..693D, 2009ApJ...695L..97C, 2011ApJ...740L..45C}.
Moreover, most GCs in the Milky Way host multiple stellar populations, and stellar evolution theory suggests that second-generation (SG) stars with enhanced helium abundance would be placed on the bluer HB in the CMD \citep{2005ApJ...621L..57L, 2008MNRAS.390..693D, 2013AJ....145...25K, 2013ApJ...762...36J}.
This is directly confirmed by spectroscopic observations, which also show that the helium rich SG stars (including BHB stars) are enhanced in nitrogen and sodium \citep[][]{2011A&A...534A.123G, 2012A&A...547C...2G, 2013A&A...549A..41G, 2013ApJ...768...27M, 2014MNRAS.437.1609M, 2012ApJ...748...62V, 2015ApJS..216...19L}.

It is now well established, however, that the formation of SG stars, with substantially enhanced helium abundance, in GCs requires special conditions in the central regions of massive proto-GCs \citep[see, e..g.,][]{2006ApJ...637L.109B, 2008MNRAS.391..825D}.
Therefore, when these GCs were disrupted to contribute stars to halo field, the preferential removal of the first-generation (FG) stars would lead to the scarcity of SG stars in the halo, which is also supported by the observations of sodium and nitrogen enhanced stars in the halo \citep{2008MNRAS.391..825D, 2009A&A...505..117C, 2011A&A...534A.136M}.
This would directly impact on the number counts employed by \citet{2015MNRAS.448L..77D} because, as described above, the helium enhanced SG populations are the most likely channel to produce BHB stars at the mean metallicity of outer halo fields \citep[${\rm \left< [Fe/H] \right> = -1.5}$; see e.g.,][]{2013ApJ...763...65A}.

In order to illustrate this effect on the BHB census of \citet{2015MNRAS.448L..77D}, in Figure~\ref{f1}, we present population models for GCs with and without SG population.
If FG stars were preferentially provided to the halo field as discussed above, the outer halo field would be similar to our model without SG stars. 
We set the metallicity of both models\footnote{The massloss efficiency value of $\eta$ is taken from \citet{2013ApJ...762...36J}. The other parameters and ingredients for the model is the same as the models described in \citet{2013ApJ...769L...3C, 2013ApJS..204....3C, 2011ApJ...740L..45C}} at ${\rm \left< [Fe/H] \right> = -1.5}$.
The age is assumed to be 11~Gyr because outer halo GCs appear to be 1 $\sim$ 2~Gyr younger compared to the inner halo GCs \citep{1994ApJ...423..248L, 2011ApJ...738...74D, 2013ApJS..204....3C}.
The helium abundance of the SG population is adopted to be ${\rm Y=0.28}$, which is required to reproduce the BHB stars in a typical outer halo GC with SG stars.
Our model with SG population is a simplified version of the model for M3 shown in \citet{2014MNRAS.443L..15J}, where FG is further divided into two subpopulations. 
Our model without SG population mimics outer halo GCs such as Ruprecht 106 and PYXIS \citep{2011ApJ...738...74D, 2013ApJ...778..186V}.
The BHB selection box is taken from Figure~1 of \citet{2015MNRAS.448L..77D}.

As is clear from Figure~\ref{f1}, the number of BHB stars abruptly increases with the presence of SG stars.
Out of 200 model HB stars in each CMD, 41 and 4 stars are placed inside the BHB selection box in the models with and without SG stars, respectively.
Adopting the number ratio between BS and all HB stars to be ${\rm N_{BS}/N_{HB}} \sim 0.3$, which is the typical value of GCs with ${\rm M_V \sim -8.0}$ \citep[see Figure~2 of][]{2007A&A...468..973M}, the number ratio between BS and BHB stars (${\rm N_{BS}/N_{BHB}}$ ) in our models ranges from 1.46 to 15.00 depending on the presence or absence of SG stars.
Therefore, the high ${\rm N_{BS}/N_{BHB}}$ ratio (from 4.9 to 6.4) observed in the outer halo field is equally well reproduced by a small number of BHB stars in our model.

\begin{table*}
  
  \caption{Selected GCs in the outer halo without SG.}
\begin{tabular}{lcccc}
    Globular Cluster & ${\rm R_{GC}}$~(kpc) & ${\rm [Fe/H]}$ & ${\rm N_{BS}/N_{BHB}}$ & CMD from \\ \hline
    NGC~362 & 9.4 & -1.3 & 3.00 & \citet{2007AaA...463..589S} \\
    Terzan~7 & 15.6 & -1.0 & 9.00 & \citet{1995AJ....109..663B} \\
    Palomar~12 & 15.8 & -1.4 & $\infty$ & \citet{1998AaA...339...61R} \\
    Ruprecht~106 & 18.5 & -1.3 & $\infty$ & \citet{2011ApJ...738...74D} \\
    Palomar~13 & 26.9 & -1.5 & $\infty$ & \citet{2011ApJ...743..167B} \\
    Pyxis & 41.4 & -1.3 & $\infty$ & \citet{2011ApJ...738...74D} \\
  \end{tabular}
  \label{t1}
\end{table*}

{Furthermore, as presented in Figure~\ref{f2}, there are sufficient number of outer halo GCs without SG stars.
In order to apply the selection box in Figure~\ref{f1} to other colors, we choose A-type stars in the same temperature range of $-0.25 \leq (g-r)_0 \leq 0.0$ \citep{2015MNRAS.448L..77D}, and then set the ranges of different colors which are consistent with the $(g-r)_0$.
The corresponding color ranges for $-0.25 \leq (g-r)_0 \leq 0.0$ are roughly $-0.10 \leq (V-I)_0 \leq 0.24$, $-0.080 \leq (B-V)_0 \leq 0.275$, and $0.05 \leq (F606W-F814W)_0 \leq 0.25$. 
After that we perform the isochrone fitting for each cluster to place BHB and BS stars in the selection box accurately.
For the comparison with observed HB stars, we also provide the mean HB position that corresponds to three different age isochrones.
Table~\ref{t1} summarizes the adopted metallicity of isochrones and the resulting ${\rm N_{BS}/N_{BHB}}$ ratio.

As is evident from this analysis, these outer halo GCs show very high ${\rm N_{BS}/N_{BHB}}$ ratios.
Both our models and observations suggest that small number of BHB stars are the main driver of the high ${\rm N_{BS}/N_{BHB}}$ ratios.
Interestingly, the ${\rm N_{BS}/N_{BHB}}$ ratio of NGC~362 and Terzan~7 is quite comparable with dwarf spheroidals ranging from 2 to 40 and the outer halo field ranging from 5 to 6 \citep{2015MNRAS.448L..77D}.
Furthermore, no presence of BHB stars in Palomar~12, Ruprecht~106, Palomar~13, and Pyxis would increase the ${\rm N_{BS}/N_{BHB}}$ ratio of each GC to the infinity.
{These GCs without BHB stars were not included in the analysis of \citet{2015MNRAS.448L..77D} because of their selection criteria (${\rm N_{BS}>1}$ and ${\rm N_{BHB}>1}$). 
Therefore, their analysis was further biased because of these selection criteria, which had led them to select against these GCs with only FG.
This bias would have also resulted in smaller ${\rm N_{BS}/N_{BHB}}$ ratio for GCs in their analysis. 
Dwarf spheroidal galaxies, however, were not affected by their selection criteria because most of them contain at least some BHB stars in the selection box. 
Note further that the highest ${\rm N_{BS}/N_{BHB}}$ ratios of Leo II ($10.0 \pm 1.0$), Cetus ($45.4 \pm 11.5$), and Sagittarius ($10.0 \pm 1.9$) reported by \citet{2015MNRAS.448L..77D} are most likely affected by younger main sequence stars reported in these stellar systems \citep{2014ApJ...789..147W}.
These illustrate how the ${\rm N_{BS}/N_{BHB}}$ ratio can be extremely sensitive to the number of BHB stars or young main sequence stars, in addition to BS stars, in GCs and dwarf spheroidal galaxies.}
}

\section{Preferential disruption of FG stars}

\begin{figure}
\includegraphics[angle=-90,scale=0.34]{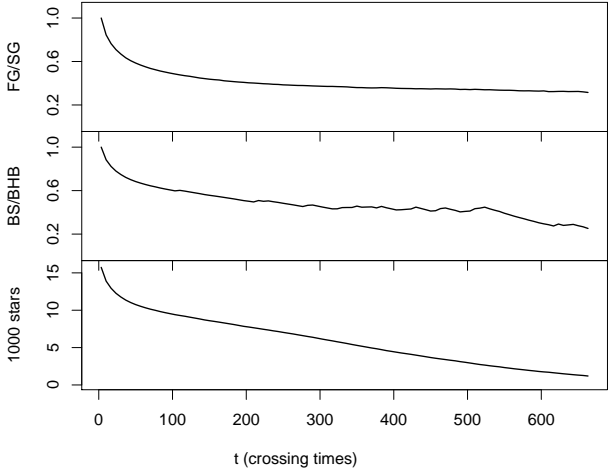}
\caption[]{
{Evolution of a simulated GC with initial conditions $W_0 = 8$, $f = 0.25$, and $\alpha = 9/16$ (see text). 
The top panel shows the evolution of $\phi$=FG/SG over time, and the middle panel shows the ${\rm N_{BS}/N_{BHB}}$ ratio of the GC.
{Both panels are normalized to the initial values.} 
The time is measured in units of the initial crossing time. 
A monotonically decreasing behavior is observed, with $\phi$ nearing {0.3} at the end of the simulation. 
This illustrates a preferential loss of FG stars. 
The bottom panel shows the total number of stars (FG + SG) in the simulation as a function of time.}
}
\label{f3}
\end{figure}

{As discussed above, our argument holds as long as the preferential removal of FG stars takes place during the early dynamical evolution of GCs, resulting in enhanced FG numbers in the halo.
The favored physical mechanism for preferential FG removal is tidal stripping after rapid initial stellar mass-loss due to FG supernovae, as discussed by \cite{2008MNRAS.391..825D}. 
We run a set of direct N-body simulations using the direct summation code NBODY6 \citep[][]{2003gnbs.book.....A} to illustrate preferential removal of FG stars in GCs. 
We considered tidally filling GCs with non-virialized initial conditions, to represent the effect of impulsive supernova mass-loss. 
The initial conditions were spherically symmetric models containing $N = 16000$ particles. We obtained them by generating King models of different initial central dimensionless potentials $2 \leq W_0 \leq 10$ (which are by construction virialized) with initially $M > N$ particles and eliminating $M - N$ particles at random. 
We later assigned particle masses proportional to present-day main-sequence mass-distributions for FG and SG stars based on our model shown in the left panel of Figure 1. The most central particles (i.e. the nearest to the model's center of mass) were assigned to represent SG stars until the desired initial fraction of SG stars $f = N_{SG}/N$ was reached, and the rest were assigned to FG stars. 
This way the initial radial distribution of the SG/FG fraction is a step function. 
While this is not realistic, it does represent the fact that SG stars are initially more concentrated than FG stars. 
In this paper we prefer to stick to a simplified model because we want to show the universal occurrence of enhanced FG star-loss, within the picture described by \citet{2008MNRAS.391..825D}. 
Including a realistic initial radial distribution of SG stars would result in a larger parameter space. 
By construction, the virial ratio obtained with our setup is $\alpha = T/|U| = M/2N$. 
We run $27$ simulations from initial conditions spanning a grid over the three-dimensional parameter space $(W_0, f, \alpha)$. 
Each simulation was run for at least $300$ crossing times (typically $\sim$$10^5$ years), and the evolution of the FG/SG star ratio $\phi$ over time was recorded, as well as the overall number of stars lost.  
The typical evolution of a simulation is shown in Figure~\ref{f3}, which refers to a model with $W_0 = 8$, $f = 0.25$, and $\alpha = 9/16$. 
In this simulation we labeled $1\%$ of the stars (both in FG and in SG) as HB stars. 
We assigned them the mass of the turn-off for the respective generation. Of these HB stars, all SG HBs are tagged BHBs.
We randomly selected 0.5\% of the stars (from both generations) and tagged them as BS. Under these assumptions, the temporal evolution of the ${\rm N_{BS}/N_{BHB}}$ ratio in our simulated cluster (top panel, Figure~\ref{f3}) closely follows that of the FG/SG ratio, showing that the cluster is depleted, over time, of FG stars, while the halo is enriched in them.

Over our sample of simulations we find that all run lose preferentially more FG stars than SG stars, i.e. $\phi$ decreases monotonically (despite small fluctuations) from its initial value $\phi_0 = (1 - f)/f$. 
The initial loss of stars is very pronounced and fast (happening within few crossing times). The unvirialized initial conditions result in the immediate expulsion of some stars from the model (those whose velocity happens to exceed the new escape velocity), but these are not preferentially FG stars. 
The later, slower mass-loss, that happens over tens to hundreds of crossing times (i.e. still quickly with respect to the GC relaxation time), is instead affecting FG stars preferentially and is due to the expansion of the GC outside of its Roche lobe. This does not happen in isolated simulations, which we run as a check and show that FG and SG stars are removed equally. 
The reason why FG stars are preferentially removed by the combination of expansion and tidal stripping is simply that they are located prevalently in the external regions of the cluster, where they are more likely to be stripped away. Dynamical relaxation is not fast enough to mix the FG and SG populations before this effect takes place, because it acts on the relaxation timescale which is orders of magnitude larger than the crossing time.

Based on our simulations we are in the position to qualitatively describe the evolution of the ${\rm N_{BS}/N_{BHB}}$ ratio both in GCs and in the outer halo field, and find it in general accord with the observations.
{A quantitative prediction of the ${\rm N_{BS}/N_{BHB}}$ ratio is beyond the scope of this letter, because a detailed understanding of the formation and evolution of BS stars would be needed. 
Nevertheless, all of our simulations, which are normalized to the initial values, clearly show a preferential removal of FG stars, which produces a decrease of the ${\rm N_{BS}/N_{BHB}}$ ratio in GCs, which in turn would increase this ratio in the halo field.
This is because BSs are most likely present in both generations, while BHBs are present only in the SG, which is mostly left in the GCs, so the ratio decreases in our simulations.
For the outer halo field, immediate comparison between the fractions of FG and SG stars in the ejecta from the simulations and observations is more complicated, because the existence of GCs without a SG should be further taken into account.}}

\section{Discussion}

Considering {various possibilities that decrease the number of BHB stars in the outer halo}, the high ${\rm N_{BS}/N_{BHB}}$ ratio observed in the outer halo can be obtained either by increasing the number of BS stars \citep[as suggested by][]{2015MNRAS.448L..77D} or by decreasing the number of BHB stars as demonstrated in our models.
Therefore, ${\rm N_{BS}/N_{BHB}}$ ratio does not provide a strong constraint as to the mass of dominant building blocks of the Milky Way, because outer halo field stars provided by GCs could have equally high ${\rm N_{BS}/N_{BHB}}$ ratio.
In addition, the apparent number of stars in the BS region of dwarf spheroidal galaxies can also be increased by the contamination from younger main-sequence stars \citep{2008A&A...482..777C}.
This can also result in the increased ratio of ${\rm N_{BS}/N_{BHB}}$.
Furthermore, the distribution and observed number of BS stars are more affected by dynamical effects, such as mass-segregation \citep[see e.g.][]{2012Natur.492..393F}, and therefore they are probably not an ideal indicator of stellar populations in GCs.
Several lines of evidence still suggest that accretion of GCs played an important role in building up the outer halo of the Milky Way.
For example, {as already pointed out by \citet{2015MNRAS.448L..77D}}, the enhanced ${\rm [\alpha/Fe]}$ of halo stars which is not observed in dwarf galaxies \citep[e.g.,][]{2003AJ....125..707T}, and the presence of two Oosterhoff groups of RR Lyrae stars both in GCs and halo fields \citep{2013AJ....146...21S, 2015arXiv150504791J} are all consistent with a common origin of halo fields and GCs in terms of stellar populations.

\section*{Acknowledgments}
C.C. acknowledges support from the Research Fellow Program (NRF-2013R1A1A2006053) of the National Research Foundation of Korea.
Y.W.L. acknowledges support from the National Research Foundation of Korea to the Center for Galaxy Evolution Research.
This work was partially supported by the KASI-Yonsei Joint Research Program (2012-2013) for the Frontiers of Astronomy and Space Science, and by the Korea Astronomy and Space Science Institute under the R\&D program (No. 2014-1-600-05) supervised by the Ministry of Science, ICT and future Planning.
M.P. acknowledges support from Mid-career Researcher Program (No. 2015-008049) through the National Research Foundation (NRF) of Korea.

\bsp

\label{lastpage}

\end{document}